\documentclass{rspublic}
\usepackage{epsf,epsfig}
\newcount\timecount
\newcount\hours \newcount\minutes  \newcount\temp \newcount\pmhours
\hours = \time
\divide\hours by 60
\temp = \hours
\multiply\temp by 60
\minutes = \time
\advance\minutes by -\temp
\def\hour{\the\hours}
\def\minute{\ifnum\minutes<10 0\the\minutes
            \else\the\minutes\fi}
\def\clock{
\ifnum\hours=0 12:\minute\ AM
\else\ifnum\hours<12 \hour:\minute\ AM
      \else\ifnum\hours=12 12:\minute\ PM
            \else\ifnum\hours>12
                 \pmhours=\hours
                 \advance\pmhours by -12
                 \the\pmhours:\minute\ PM
                 \fi
            \fi
      \fi
\fi
}

\def\monthname{\relax\ifcase\month 0/\or January\or February\or
   March\or April\or May\or June\or July\or August\or September\or
   October\or November\or December\else\number\month/\fi}

\def\bold#1{\setbox0=\hbox{$#1$}%
     \kern-.025em\copy0\kern-\wd0
     \kern.05em\copy0\kern-\wd0
     \kern-.025em\raise.0433em\box0 }

\def\beq{\begin{equation}}
\def\eeq{\end{equation}}

\def\ga{\mathrel{\raise.3ex\hbox{$>$\kern-.75em\lower1ex\hbox{$\sim$}}}}
\def\la{\mathrel{\raise.3ex\hbox{$<$\kern-.75em\lower1ex\hbox{$\sim$}}}}
\def\gev{{\rm \, Ge\kern-0.125em V}}
\def\tev{{\rm \, Te\kern-0.125em V}}
\def\gyr{{\rm \, G\kern-0.125em yr}}

\def\gappeq{\mathrel{\rlap {\raise.5ex\hbox{$>$}}
{\lower.5ex\hbox{$\sim$}}}}
\def\lappeq{\mathrel{\rlap{\raise.5ex\hbox{$<$}}
{\lower.5ex\hbox{$\sim$}}}}
\def\Toprel#1\over#2{\mathrel{\mathop{#2}\limits^{#1}}}

\def\sl{{\widetilde \ell}_{\scriptscriptstyle\rm R}}

\def\m12{m_{1\!/2}}
\begin{document}
\title[Dark Matter and Dark Energy]{\bf Dark Matter and Dark Energy: \\
Summary and Future Directions}
\author[J. Ellis]{John Ellis}
\affiliation{TH Division, CERN, Geneva, Switzerland\\[2ex]
\rm CERN--TH/2003-086 \quad astro-ph/0304183 \quad April 
2003}

\maketitle
\begin{abstract}{cosmology, particle physics, dark matter, dark energy}
\noindent
This paper reviews the progress reported at this Royal Society Discussion
Meeting and advertizes some possible future directions in our drive to
understand dark matter and dark energy. Additionally, a first attempt is
made to place in context the exciting new results from the WMAP satellite,
which were published shortly after this Meeting. In the first part of this
review, pieces of observational evidence shown here that bear on the
amounts of dark matter and dark energy are reviewed. Subsequently,
particle candidates for dark matter are mentioned, and detection
strategies are discussed. Finally, ideas are presented for calculating the
amounts of dark matter and dark energy, and possibly relating them to
laboratory data.
\end{abstract}

\section{The Density Budget of the Universe}
\label{section1}

It is convenient to express the mean densities $\rho_i$ of various
quantities in the Universe in terms of their fractions relative to the
critical density: $\Omega_i \equiv \rho_i / \rho_{crit}$. The theory of
cosmological inflation strongly suggests that the the total density should
be very close to the critical one: $\Omega_{tot} \simeq 1$, and this is
supported by the available data on the cosmic microwave background
radiation (CMB) (Bond 2003). The fluctuations observed in the CMB at a
level $\sim 10^{-5}$ in amplitude exhibit a peak at a partial wave $\ell
\sim 200$, as would be produced by acoustic oscillations in a flat
Universe with $\Omega_{tot} \simeq 1$. At
lower partial waves, $\ell \ll 200$, the CMB fluctuations are believed to
be dominated by the Sachs-Wolfe effect due to the gravitational potential,
and more acoustic oscillations are expected at larger $\ell > 200$, whose
relative heights depend on the baryon density $\Omega_b$. At even larger
$\ell \gappeq 1000$, these oscillations should be progressively damped 
away.

\begin{figure}[t]
\begin{center}
\vspace*{-20mm}
\hspace*{-10mm}
\includegraphics[angle=90,width=0.7\textwidth]{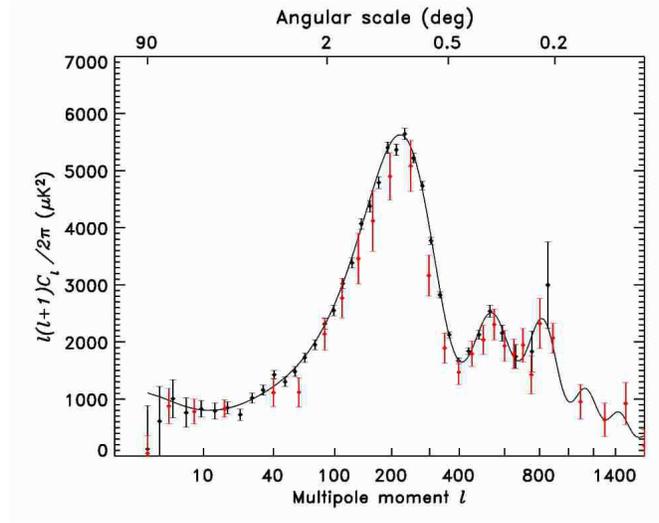}
\end{center}
\caption{\it
Spectrum of fluctuations in the cosmic microwave background measured by 
WMAP (darker points with smaller error bars), compared with previous 
measurements (lighter points with larger error bars, 
extending to greater $\ell$) (Hinshaw {\it et al.} 2003).} 
\vspace*{0.5cm}
\label{fig:RS1}
\end{figure}

\begin{figure}[htb]
\centerline{\epsfxsize = 0.7\textwidth \epsffile{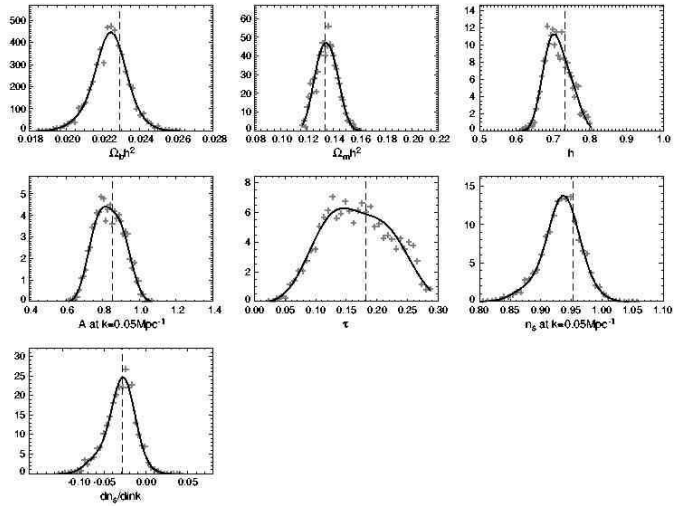}
}
\caption{\it
The likelihood functions for various cosmological parameters obtained 
from the WMAP data analysis (Spergel {\it et al.} 2003). The panels show 
the baryon 
density $\Omega_b h^2$, the matter density $\Omega_m h^2$, the Hubble 
expansion rate $h$, the strength $A$, the 
optical depth $\tau$, the spectral index $n_s$ and its rate of 
change $d n_s / {\rm ln}k$, respectively.} 
\vspace*{0.5cm} 
\label{fig:RS3} 
\end{figure}

Fig.~\ref{fig:RS1} compares measurements of CMB fluctuations made before
WMAP (Bond 2003) with the WMAP data themselves (Bennett {\it et al.} 2003;
Hinshaw {\it et al.} 2003), that were released shortly after this Meeting.
The position of the first acoustic peak indeed corresponds to a flat
Universe with $\Omega_{tot} \simeq 1$: in particular, now WMAP finds
$\Omega_{tot} = 1.02 \pm 0.02$ (Spergel {\it et al.} 2003), and two more
acoustic peaks are established with high significance, providing a new
determination of $\Omega_b h^2 = 0.0224 \pm 0.0009$, where $h \sim 0.7$
is the present Hubble expansion rate $H$, measured in units of
$100$km/s/Mpc. The likelihood functions for various cosmological
parameters are shown in Fig.~\ref{fig:RS3}. Remarkably, there is excellent
consistency between the estimate of the present-day Hubble constant $H
\sim 72$km/s/Mpc from WMAP (Spergel {\it et al.} 2003) with that inferred
from the local distance ladder based, e.g., on Cepheid variables.

\begin{figure}[htb]
\centerline{\epsfxsize = 0.8\textwidth \epsffile{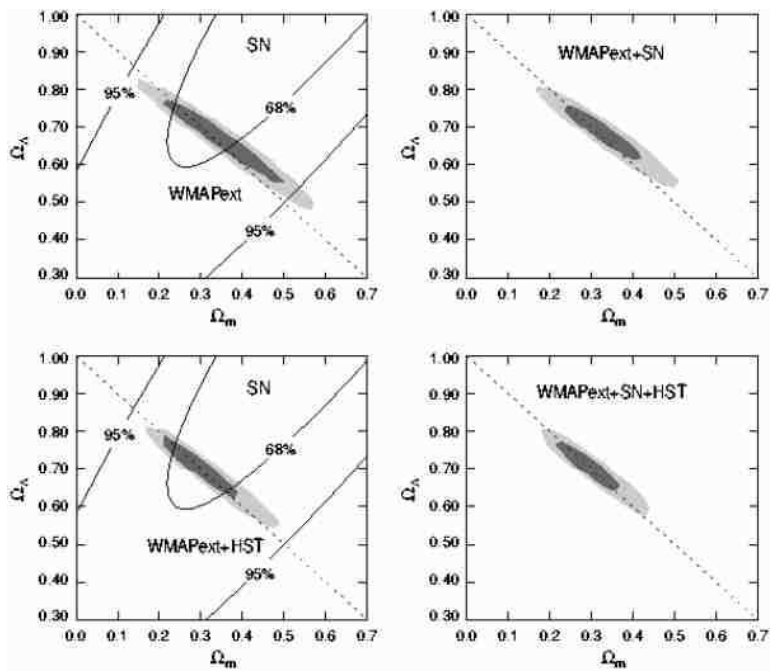}}
\caption{\it
The density of matter $\Omega_m$ and dark energy $\Omega_\Lambda$ 
inferred from WMAP and other CMB data (WMAPext), and from combining them 
with supernova and Hubble Space Telescope data (Spergel {\it et al.} 
2003).} 
\vspace*{0.5cm}
\label{fig:RS6}
\end{figure}  

As seen in Fig~\ref{fig:RS6}, the combination of CMB data with those on
high-redshift Type-Ia supernovae (Perlmutter 2003; Perlmutter \& Schmidt 
2003) and on large-scale
structure (Peacock 2003$a$, $b$) favour strongly a 
flat
Universe with about 30 \% of (mainly dark) matter and 70 \% of vacuum
(dark) energy. Type-Ia supernovae probe the geometry of the Universe at
redshifts $z \lappeq 1$. They disagree with a flat $\Omega_{tot} = 1$
Universe that has no vacuum energy, and also with an open $\Omega_m \simeq
0.3$ Universe (Perlmutter 2003; Perlmutter \& Schmidt 2003).  They appear 
to be adequate standard
candles, and two observed supernovae with $z > 1$ argue strongly against
dust or evolution effects that would be sufficient to cloud their
geometrical interpretation. The supernovae indicate that the expansion of
the Universe is currently accelerating, though it had been decelerating
when $z$ was $> 1$. There are good prospects for improving substantially
the accuracy of the supernova data, by a combination of continued
ground-based and subsequent space observations using the SNAP satellite
project (Perlmutter 2003; Perlmutter \& Schmidt 2003).

\begin{figure}[htb]
\centerline{\epsfxsize = 0.5\textwidth \epsffile{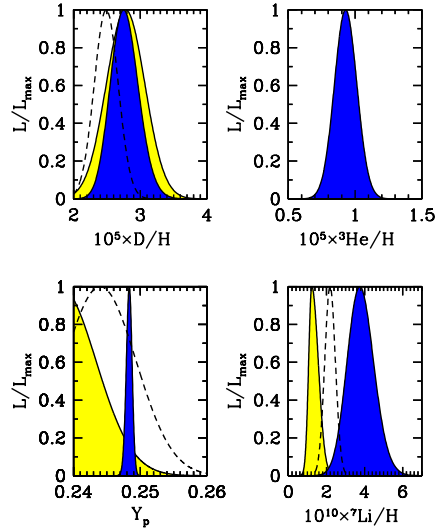}}
\caption{\it
The likelihood functions for the primordial abundances of light elements 
inferred from astrophysical observations (lighter, yellow shaded regions)
compared with those calculated using the CMB value of $\Omega_b h^2$ 
(darker, blue shaded regions). The dashed curves are likelihood functions 
obtained under different astrophysical assumptions (Cyburt {\it et al.}
2003).} 
\vspace*{0.5cm}
\label{fig:RS7}
\end{figure}

It is impressive that the baryon density inferred from WMAP
data (Spergel {\it et al.} 2003) is in good agreement with the value 
calculated previously
on the basis of Big-Bang nucleosynthesis (BBN), which depends on
completely different (nuclear)  physics.  Fig.~\ref{fig:RS7} compares the
abundances of light elements calculated using the WMAP value of $\Omega_b
h^2$ with those inferred from astrophysical data (Cyburt {\it et al.} 
2003). Depending on
the astrophysical assumptions that are made in extracting the
light-element abundances from astrophysical data, there is respectable
overlap.

As we heard at this Meeting, several pillars of inflation theory have now 
been verified by WMAP and other CMB data (Bond 2003): the Sachs-Wolfe 
effect due to fluctuations in the
large-scale gravitational potential were first seen by the COBE satellite,
the first acoustic peak was seen in the CMB spectrum at $\ell \sim 210$
and this has been followed by two more peaks and the intervening dips, the
damping tail of the fluctuation spectrum expected at $\ell \gappeq 1000$ 
has been
seen, polarization has been observed, and the primary anisotropies are
predominantly Gaussian. WMAP has, additionally, measured the thickness of
the last scattering surface and observed the reionization of the Universe
when $z \sim 20$ by the first generation of stars (Kogut {\it et al.} 
2003). Remaining
to be established are secondary anisotropies, due, e.g., to the
Sunyaev-Zeldovich effect, weak lensing and inhomogeneous reionization, and
tensor perturbations induced by gravity waves.

As we also heard at this meeting, the values of $\Omega_{CDM}$ inferred
from X-ray studies of gas in rich clusters using the Chandra satellite
(Rees 2003), which indicate $\Omega_{CDM} = 0.325 \pm 0.34$, gravitational
lensing (Schneider 2003) and data on large-scale structure, e.g., from the
2dF galaxy redshift survey (Peacock 2003$a$, $b$), are very consistent with
that inferred by combining CMB and supernova data. The WMAP data confirm
this concordance with higher precision: $\Omega_{CDM} h^2 = 0.111 \pm
0.009$ (Spergel {\it et al.} 2003).

\begin{figure}[htb]
\hspace*{-10mm}
\centerline{\epsfxsize = 0.7\textwidth \epsffile{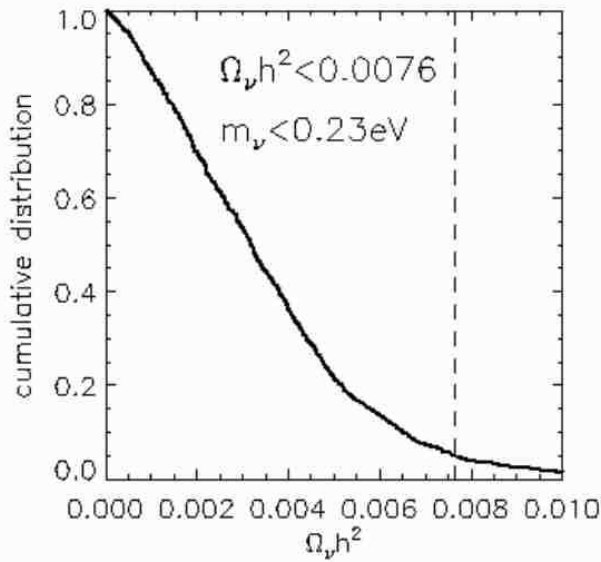}}
\caption{\it
The likelihood function for the total neutrino density $\Omega_\nu h^2$ 
derived by WMAP (Spergel {\it et al.} 2003). The upper limit $m_\nu < 
0.23$~eV applies if there are three degenerate neutrinos.}
\vspace*{0.5cm}
\label{fig:RS5}
\end{figure}

The 2dF galaxy survey has examined two wedges through the Universe.
Significant structures are seen at low redshifts, which die
away at larger redshifts where the Universe becomes more homogeneous and
isotropic. The perturbation power spectrum at these large scales matches
nicely with that seen in the CMB data, whilst the structures seen at small
scales would {\it not} be present in a baryon-dominated Universe, or one 
with a
significant fraction of hot dark matter. Indeed, the 2dF data were
used to infer an upper limit on the sum of the neutrino masses of 
1.8~eV (Elgaroy {\it et al.} 2002),
which has recently been improved using WMAP data (Spergel {\it et al.} 
2003) to
\begin{equation}
\Sigma_{\nu_i} m_{\nu_i} < 0.7~{\rm eV},
\label{cosmonu}
\end{equation}
as seen in Fig.~\ref{fig:RS5}.
This impressive upper limit is substantially better than
even the most stringent direct laboratory upper limit on an individual
neutrino mass, as discussed in the next Section. Thw WMAP data also 
provide (Crotty {\it et al.} 2003) a new limit on the effective number of 
light neutrino species, beyond the three within the Standard Model:
\begin{equation}
-1.5 \; < \; \Delta N^{eff}_\nu \; < \; 4.2.
\label{nnu}
\end{equation}
This limit is not as stringent as that from LEP, but applies to additional 
light degrees of freedom that might not be produced in $Z$ decay.

\section{What is it?}

As discussed here by Kolb (2003), particle candidates for dark matter
range from the axion with a mass $\gappeq 10^{-15}$~GeV (van Bibber 2003)
to cryptons with masses $\lappeq 10^{+15}$~GeV (Ellis {\it et al.} 1990; 
Benakli {\it et al.} 1999), via
neutrinos with masses $\lappeq 10^{-10}$~GeV, the gravitino and the
lightest supersymmetric particle with a mass $\gappeq
10^{2}$~GeV (Ellis {\it et al.} 1984; Goldberg 1983). In recent years, 
there has been considerable
experimental progress in understanding neutrino masses, so I start with
them, even though cosmology now disfavours the hot dark matter they would
provide (Spergel {\it et al.} 2003). All the others are candidates for 
cold dark matter,
except for the gravitino, which might constitute warm dark matter, another
possibility now disfavoured by the WMAP evidence for reionization when $z
\sim 20$ (Kogut {\it et al.} 2003).

\subsection{Neutrinos}

Particle theorists expect particles to have masses that vanish exactly 
only if they are protected by some unbroken gauge symmetry, much as the 
photon is massless because of the U(1) gauge symmetry of electromagnetism, 
that is associated with the conservation of electric charge. There is no 
corresponding exact gauge symmetry to protect lepton number, so we expect 
it to be violated and neutrinos to acquire masses. This is indeed the 
accepted interpretation of the observed oscillations between different 
types of neutrinos, which are made possible by mixing into non-degenerate 
mass eigenstates (Wark 2003); Pakvasa \& Valle 2003).

Neutrino masses could arise even within the Standard Model of particle 
physics, without adding any new particles, at the expense of introducing a 
interaction between two neutrino fields and two Higgs 
fields (Barbieri {\it et al.} 1980):
\begin{equation}
{1 \over M} \nu H \cdot \nu H \; \rightarrow \; m_\nu = {\langle 0 | H 0 
\rangle^2 \over M}.
\label{numass}
\end{equation}
However, such an interaction would be non-renormalizable, and therefore 
is not thought 
to be fundamental. The (presumably large) mass scale $M$ appearing the 
denominator of (\ref{numass}) is generally thought to originate from the 
exchange of some massive fermionic particle that mixes with the light 
neutrino (Gell-Mann {\it et al.} 1979; Yanagida 1979; Mohapatra \& 
Senjanovic 1980):
\begin{eqnarray}
\left( \nu_L, N\right) \left(
\begin{array}{cc}
0 & M_{D}\\
M_{D}^{T} & M
\end{array}
\right)
\left(
\begin{array}{c}
\nu_L \\
N
\end{array}
\right),
\label{seesaw}
\end{eqnarray}
Diagonalization of this matrix naturally yields small neutrino masses, 
since we expect that the Dirac mass term $m_D$ is of the same order as 
quark and lepton masses, and $M \gg m_W$.

We have the following direct experimental upper limits on neutrino masses. 
From measurements of the end-point in Tritium $\beta$ decay, we know 
that (Weinheimer {\it et al.} 1999; Lobashov {\it et al.} 1999): 
\begin{equation}
m_{\nu_e} \; \lappeq \; 2.5~{\rm eV},
\label{mnue}
\end{equation}
and there are prospects to improve this limit down to about 0.5~eV with 
the proposed KATRIN experiment (Osipowicz {\it et al.} 2001). From 
measurements of $\pi 
\to \mu \nu$ decay, we know that (Hagiwara {\it et al.} 2002):
\begin{equation}
m_{\nu_\mu} \; < \; 190~{\rm KeV},
\label{mnumu}
\end{equation}
and there are prospects to improve this limit by a factor $\sim 20$. 
From measurements of $\tau \to n \pi \nu$ decay, we know 
that (Hagiwara {\it et al.} 2002):
\begin{equation}
m_{\nu_\tau} \; < \; 18.2~{\rm MeV},
\label{mnutau}
\end{equation}
and there are prospects to improve this limit to $\sim 5$~MeV.

However, the most stringent laboratory limit on neutrino masses may come 
from searches for neutrinoless double-$\beta$ decay, which constrain the 
sum of the neutrinos' Majorana masses weighted by their couplings to 
electrons (Klapdor-Kleingrothaus {\it et al.} 2001):
\begin{equation}
\langle m_\nu \rangle_e \; \equiv | \Sigma_{\nu_i} m_{\nu_i} U_{ei}^2 |
\lappeq 0.35~{\rm eV}
\label{doublebeta}
\end{equation}
and there are prospects to improve this limit to $\sim 0.01$~eV in a 
future round of experiments. The impact of the limit (\ref{doublebeta}) in 
relation to the cosmological upper limit (\ref{cosmonu}) is 
discussed below, after we have gathered further experimental input from 
neutrino-oscillation experiments.

The neutrino mass matrix (\ref{seesaw}) should be regarded also as a 
matrix in flavour space. When it is diagonalized, the neutrino mass 
eigenstates will not, in general, coincide with the flavour eigenstates 
that partner the mass eigenstates of the charged leptons. The mixing 
matrix between them (Maki {\it et al.} 1962) may be written in the form
\begin{eqnarray}
V \; = \; \left(
\begin{array}{ccc}
c_{12} & s_{12} & 0 \\
- s_{12} & c_{12} & 0 \\
0 & 0 & 1
\end{array} 
\right)
\left(
\begin{array}{ccc}
1 & 0 & 0 \\
0 & c_{23} & s_{23} \\
0 & - s_{23} & c_{23}
\end{array}
\right)
\left(
\begin{array}{ccc}
c_{13} & 0 & s_{13} \\
0 & 1 & 0 \\
- s_{13} e^{- i \delta} & 0 & c_{13} e^{- i \delta}
\end{array}
\right).
\label{MNS}
\end{eqnarray}
where the symbols $c,s_{ij}$ denote the standard trigonometric functions 
of the three real `Euler' mixing angles $\theta_{12,23,31}$, and $\delta$ is a 
CP-violating phase that can in principle also be observed in 
neutrino-oscillation experiments (De R\'ujula {\it et al.} 1999). 
Additionally, there are two 
CP-violating phases $\phi_{1,2}$ that appear in the double-$\beta$ 
observable (\ref{doublebeta}), but do not affect neutrino 
oscillations.

\begin{figure}[htb]
\centerline{\epsfxsize = 0.5\textwidth \epsffile{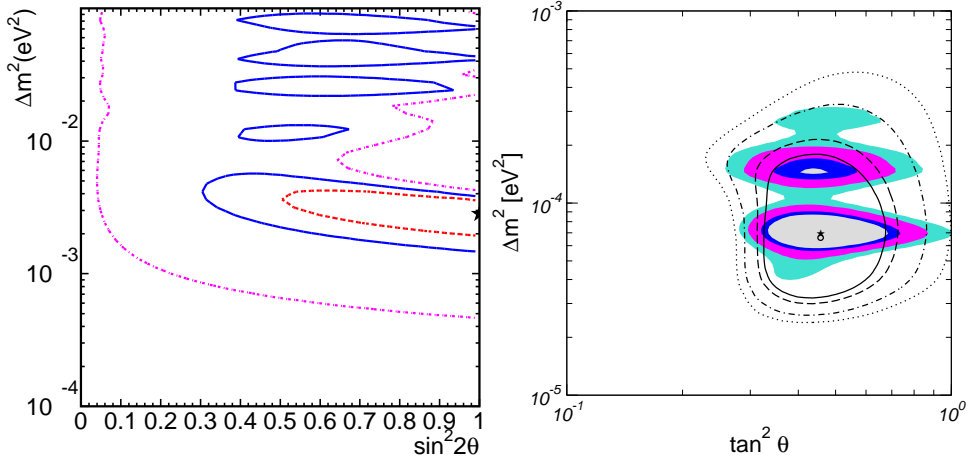}
\hfill \epsfxsize = 0.5\textwidth \epsffile{RS17.eps}
}
\caption{\it
Left panel: The region of neutrino oscillation parameters $(\sin^2 
2 \theta, \Delta
m^2)$ inferred from the K2K reactor experiment (Ahn {\it et al.} 2002) 
includes the 
central values favoured by the Super-Kamiokande atmospheric-neutrino 
experiment, indicated by the star (Fukuda {\it et al.} 1998). Right 
panel: The 
region of neutrino oscillation parameters $(\tan^2 \theta, \Delta 
m^2)$ inferred from solar-neutrino experiments is very consistent with 
derived from the KamLAND reactor neutrino experiment (Eguchi {\it et al.} 
2003). The 
shaded regions show the combined probability distribution (Pakvasa \& 
Valle, 2003).} 
\vspace*{0.5cm} \label{fig:RS17}
\end{figure}

The pioneering Super-Kamiokande and other experiments have shown that 
atmospheric neutrinos oscillate, with the following difference in squared 
masses and mixing angle (Fukuda {\it et al.} 1998):
\begin{equation}
\delta m^2 \; \simeq \; 2.4 \times 10^{-3}~{\rm eV}^2, \; \sin^2 
2 \theta_{23} \simeq 1.0,
\label{atmo}
\end{equation}
which is very consistent with the K2K reactor neutrino 
experiment (Ahn {\it et al.} 2002), as seen in the left panel of 
Fig.~\ref{fig:RS17}.
A flurry of recent solar neutrino experiments, most notably 
SNO (Ahmad {\it et al.} 2002$a$, $b$), have 
established beyond any doubt that they also oscillate, with
\begin{equation}
\delta m^2 \; \simeq \; 6 \times 10^{-5}~{\rm eV}^2, \; \tan^2
\theta_{31} \simeq 0.5.
\label{solar}
\end{equation}
Most recently, the KamLAND experiment has reported a deficit of electron 
antineutrinos from nuclear power reactors, leading to a very similar set 
of preferred parameters, as seen in the 
right panel of Fig.~\ref{fig:RS17} (Eguchi {\it et al.} 2003).

\begin{figure}[htb]
\begin{center}
\vspace*{-20mm}
\hspace*{-30mm}
\includegraphics[scale=0.4]{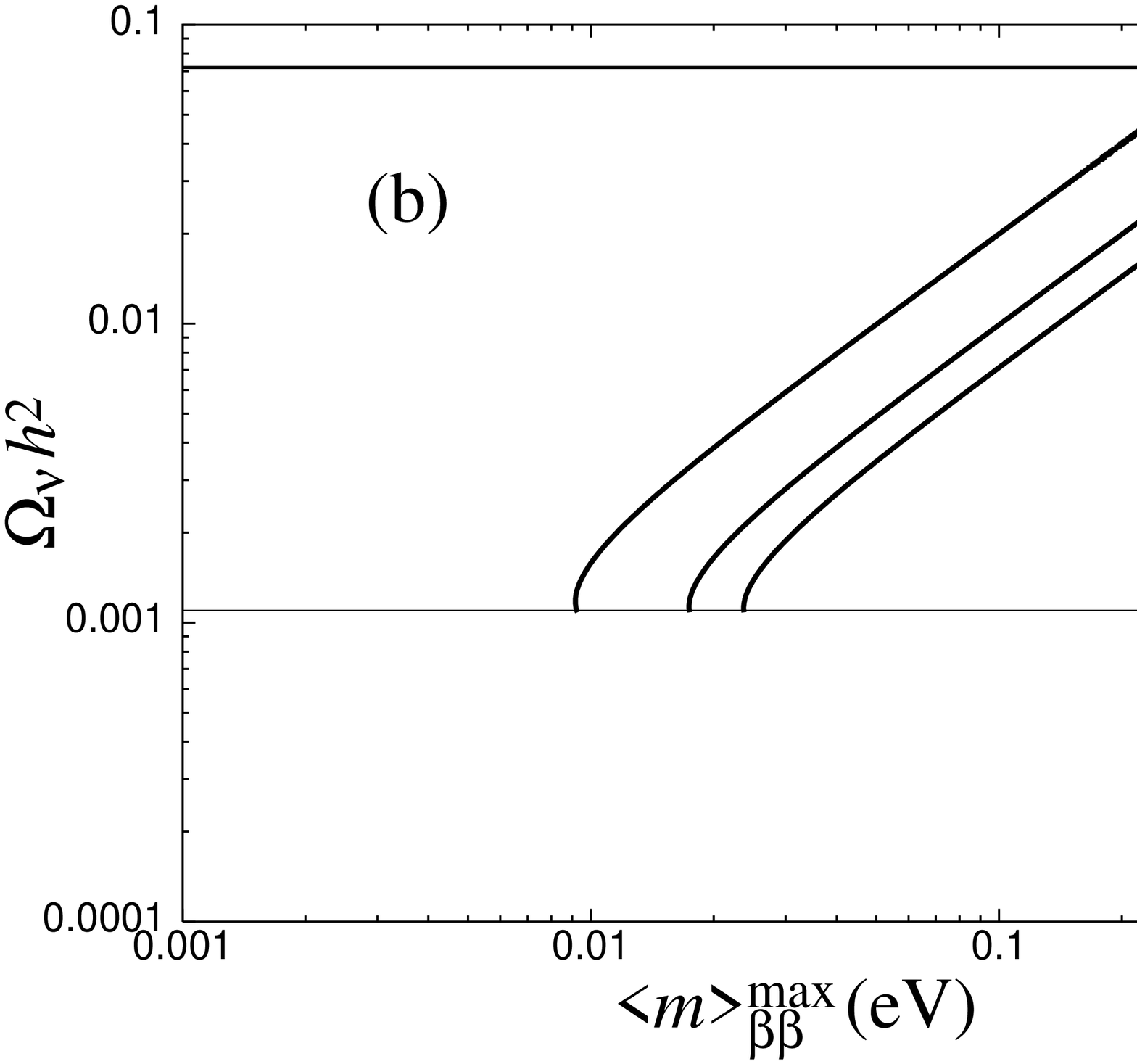}
\end{center}
\caption{\it
Correlation between $\Omega_\nu h^2$ and $\langle m_\nu \rangle_e$. The 
different diagonal lines correspond to uncertainties in the measurements 
by KamLAND and other experiments (Minakata \& Sugiyama 2002).}
\label{fig:MS}
\end{figure}

Using the range of $\theta_{12}$ allowed by the solar and KamLAND data, 
one can establish a correlation between the relic neutrino density 
$\Omega_\nu h^2$ and the neutrinoless double-$\beta$ decay observable
$\langle m_\nu \rangle_e$, as seen in Fig.~\ref{fig:MS} (Minakata \& 
Sugiyama 2002). Pre-WMAP, the 
experimental limit on $\langle m_\nu \rangle_e$ could be used to set 
the bound (Minakata \& Sugiyama 2002)
\begin{equation}
10^{-3} \; \lappeq \; \Omega_\nu h^2 \; \lappeq \; 10^{-1}.
\label{MSbound}
\end{equation}
Alternatively, now that WMAP has set a tighter upper bound $\Omega_\nu 
h^2 < 0.0076$ (\ref{cosmonu}), one can use this correlation to set an 
upper bound:
\begin{equation}
< m_\nu >_e \; \lappeq \; 0.1~{\rm eV},
\label{betabound}
\end{equation}
which is difficult to reconcile with the signal reported in 
(Klapdor-Kleingrothaus {\it et al.} 2002).

The `Holy Grail' of neutrino physics is CP violation in neutrino 
oscillations, which would manifest itself as a difference between the 
oscillation probabilities for neutrinos and antineutrinos (De R\'ujula 
{\it et al.} 1999):
\begin{eqnarray}
P \left( \nu_e \to \nu_\mu \right) - P \left( {\bar \nu}_e \to
{\bar \nu}_\mu \right) \; & & = \;
16 s_{12} c_{12} s_{13} c^2_{13} s_{23} c_{23} \sin \delta \\ \nonumber
& & \sin \left( {\Delta m_{12}^2 \over 4 E} L \right)
\sin \left( {\Delta m_{13}^2 \over 4 E} L \right)
\sin \left( {\Delta m_{23}^2 \over 4 E} L \right),
\label{CPosc}
\end{eqnarray}
For this to be observable, $\Delta m_{12}$ and $\theta_{12}$ have to be 
large, as SNO and KamLAND have shown to be the case, and also 
$\theta_{13}$ has to be large enough - which remains to be seen.

In fact, even the minimal seesaw model contains many additional parameters
that are not observable in neutrino oscillations (Casas \& Ibarra 2001). 
In addition to the three masses of the charged leptons, there are three
light-neutrino masses, three light-neutrino mixing angles and three
CP-violating phases in the light-neutrino sector: the oscillation phase
$\delta$ and the two Majorana phases that are relevant to neutrinoless
double-$\beta$ decay experiments. As well, there are three heavy
singlet-neutrino masses, three more mixing angles and three more
CP-violating phases that become observable in the heavy-neutrino sector,
making a total of 18 parameters. Out of all these parameters, so far just
four light-neutrino parameters are known, two differences in masses
squared and two real mixing angles.

As discussed later, it is often thought that there may be a connection
between CP violation in the neutrino sector and the baryon density in the
Universe, via leptogenesis. Unfortunately, this connection is somewhat
indirect, since the three CP-violating phases measurable in the
light-neutrino sector to not contribute to leptogenesis, which is
controlled by the other phases that are not observable directly at low
energies (Ellis \& Raidal 2002).  However, if the seesaw model is combined
with supersymmetry, these extra phases contribute to the renormalization
of soft superymmetry-breaking parameters at low energies, and hence may
have indirect observable effects (Ellis {\it et al.} 2002$a$, $b$).

\subsection{Problems with Cold Dark Matter?}

A recurring question is whether the cold dark matter paradigm for
structure formation is compatible with all the observational
data (Navarro 2003; Abadi {\it et al.} 2002). One of the issues is the 
mass profile at the core of
a galactic halo. As we heard at this meeting, density profiles clearly
differ from naive power laws, and are much shallower than simple
isothermal models. However, the haloes of $\gamma < 2.5$ seem to be in
reasonable agreement with CDM sumulations, though there are still problems
with $\gamma > 2.5$ galaxies.  The theoretical predictions are not yet
conclusive, though, with questions such as triaxiality, departures from
equilibrium and time dependence remaining to be resolved. Another issue is
halo substructure and the abundance of Milky Way satellites. However, the
latest news is that there is apparently better agreement between the
number of Milky Way satellites and the number of massive halo
substructures predicted in CDM simulations.  
It has been argued that a
Milky Way-like stellar disk would be thickened if there were substructures
in the halo, but recent simulations do not display any significant such
effect (Navarro 2003; Abadi {\it et al.} 2002).

Thus, the latest advice from the simulators seems to be that there is no
showstopper for the CDM paradigm, so let us examine some of the
candidates for the CDM.

\subsection{Axions}

\begin{figure}[t]
\centerline{\epsfxsize = 0.7\textwidth \epsffile{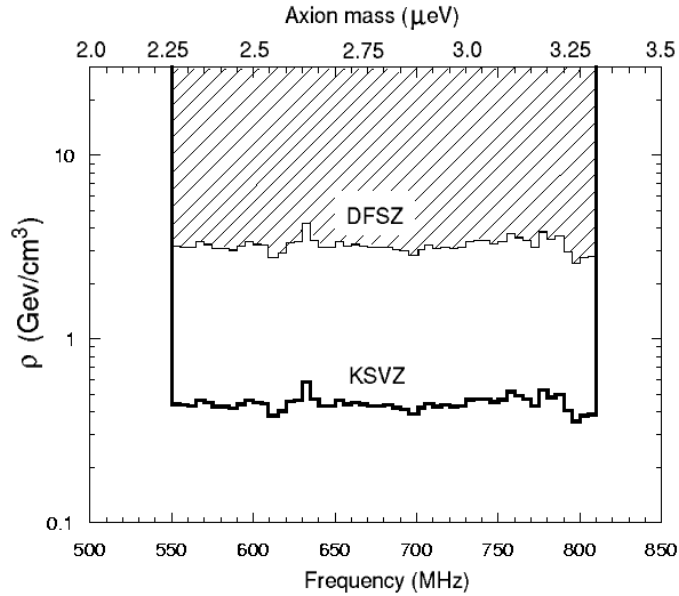}
}
\caption{\it
Exclusion domains for the halo density of different types of axion, as 
functions of the frequency corresponding to the possible axion 
mass, as obtained from microwave cavity 
experiments (van Bibber 2003; Asztalos {\it et al.} 2001).} 
\vspace*{0.5cm}
\label{fig:RS10}
\end{figure}

As we heard here from van Bibber (van Bibber 2003), axions were invented
in order to conserve CP in the strong interactions: in the picturesque
analogy of Sikivie (Sikivie 1996), to explain why the strong-interaction
pool table is apparently horizontal. Axion-like particles may be
characterized by a two-dimensional parameter space, consisting of the
axion mass $m_a$ and its coupling to pairs of photons, $g_{a\gamma}$. Many
areas of this parameter space are excluded by laser experiments,
telescopes, searches for solar axions currently being extended by the CAST
experiment at CERN (Irastorza {\it et al.} 2002), astrophysical
constraints (Hagiwara {\it et al.} 2002) and searches for halo axions
using microwave cavities (Asztalos {\it et al.} 2001), as seen in
Fig.~\ref{fig:RS10}. These and searches using Rydberg atoms (Yamamoto {\it
et al.} 2001) have the best chances of excluding ragions of parameter
space where the axion might constitute cold dark matter.

\subsection{ Supersymmetric Dark Matter}

The appearance of supersymmetry at the TeV scale was originally motivated
by the hierarchy problem (Maiani 1979; 't Hooft 1979; Witten 1981) - why 
is $m_W \ll m_P \sim 
10^{19}$~GeV, the only
candidate we have for a fundamental mass scale in physics, or
alternatively why is the Coulomb potential in an atom so much larger than
the Newton potential? The former is $\propto e^2 = {\cal O} (1)$, whereas
the latter is $\propto G_N m^2 \sim m^2 / m_P^2$, where $m$ is a typical
particle mass scale. It is not sufficient simply to set particle masses 
such as $m_W \ll m_P$, since quantum corrections will increase them again. 
Many one-loop quantum corrections each yield
\begin{equation}
\delta m_W^2 \simeq {\cal O} \left( {\alpha \over \pi} \right) \Lambda^2,
\label{quadratic}
\end{equation}
where $\Lambda$ is an effective cut-off, representing the scale at which 
the Standard Model ceases to be valid, and new physics appears. If 
$\Lambda \simeq m_P$ or the GUT scale $\sim 10^{16}$~GeV, the `small' 
correction (\ref{quadratic}) will be much larger than the physical value 
of $m_W^2$. It would require `unnatural' fine-tuning to choose the bare 
value of $m_W^2$ to be almost equal and opposite to the `small'
correction (\ref{quadratic}), so that their combination happens to have 
the right magnitude. Alternatively, supersymmetry introduces an effective 
cut-off $\Lambda$ by postulating equal numbers of bosons and fermions with 
identical couplings, in which case the corrections (\ref{quadratic}) 
cancel among themselves, leaving
\begin{equation}
\delta m_W^2 \simeq {\cal O} \left( {\alpha \over \pi} \right) ( m_B^2 - 
m_F^2 ),
\label{susy}
\end{equation}
which is $\lappeq m_W^2$ if
\begin{equation}
| m_B^2 -  m_F^2 | \lappeq {\cal O} (1)~{\rm TeV}^2,
\label{TeV}
\end{equation}
i.e., if supersymmetry appears at relatively low energy. 

This naturalness argument for low-energy supersymmetry is supported by
several pieces of indirect empirical evidence. {\it One} is that the
strengths of the different gauge interactions measured at low energies,
particularly at the LEP accelerator, are consistent with unification at
high energies {\it if} there are low-mass supersymmetric particles (Ellis
{\it et al.} 1991$b$; Amaldi {\it et al.} 1991; Langacker \& Luo 1991;
Giunti {\it et al.} 1991). {\it A second} is that LEP and other precision
low-energy data are fitted well by the Standard Model {\it if} there is a
relatively light Higgs boson with mass $< 200$~GeV, very consistent with
the range $m_h \lappeq 130$~GeV predicted by supersymmetry (Okada {\it et
al.} 1991; Ellis {\it et al.} 1991$a$; Haber \& Hempfling 1991). {\it A
third} is that the minimal supersymmetric extension of the Standard Model
(MSSM) contains a good candidate $\chi$ for cold dark matter, which has a
suitable relic density if the supersymmetric mass scale $\lappeq 1$~TeV
(Ellis {\it et al.} 1984; Goldberg 1983).  {\it A fourth} may be provided
by the anomalous magnetic moment of the muon, $g_\mu - 2$ (Brown {\it et 
al.} 2001; Bennett {\it et al.} 2002), {\it
if} its experimental value deviates significantly from the Standard Model
prediction.

\begin{figure}[htb]
\centerline{\epsfxsize = 0.5\textwidth \epsffile{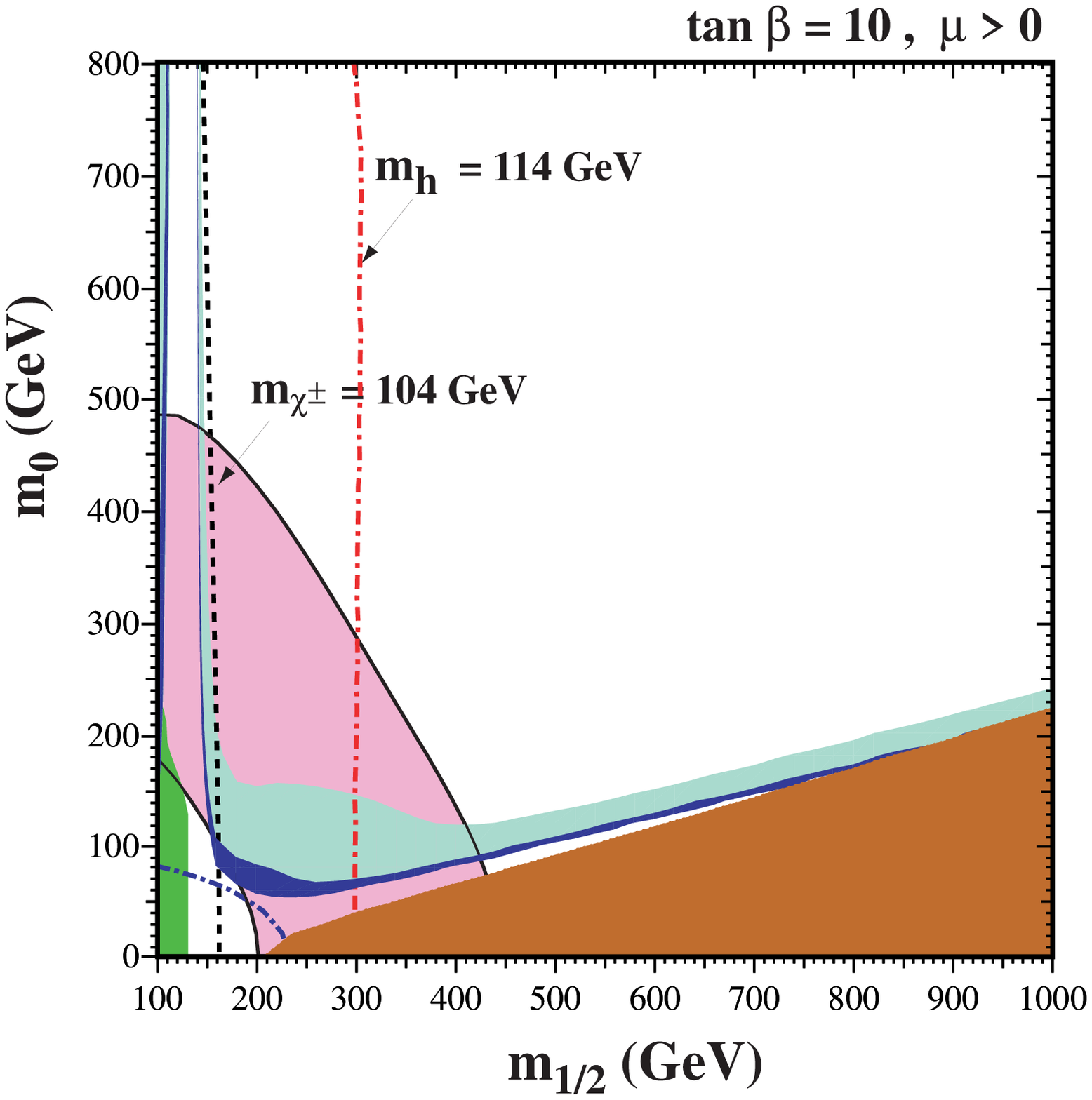}
\hfill \epsfxsize = 0.5\textwidth \epsffile{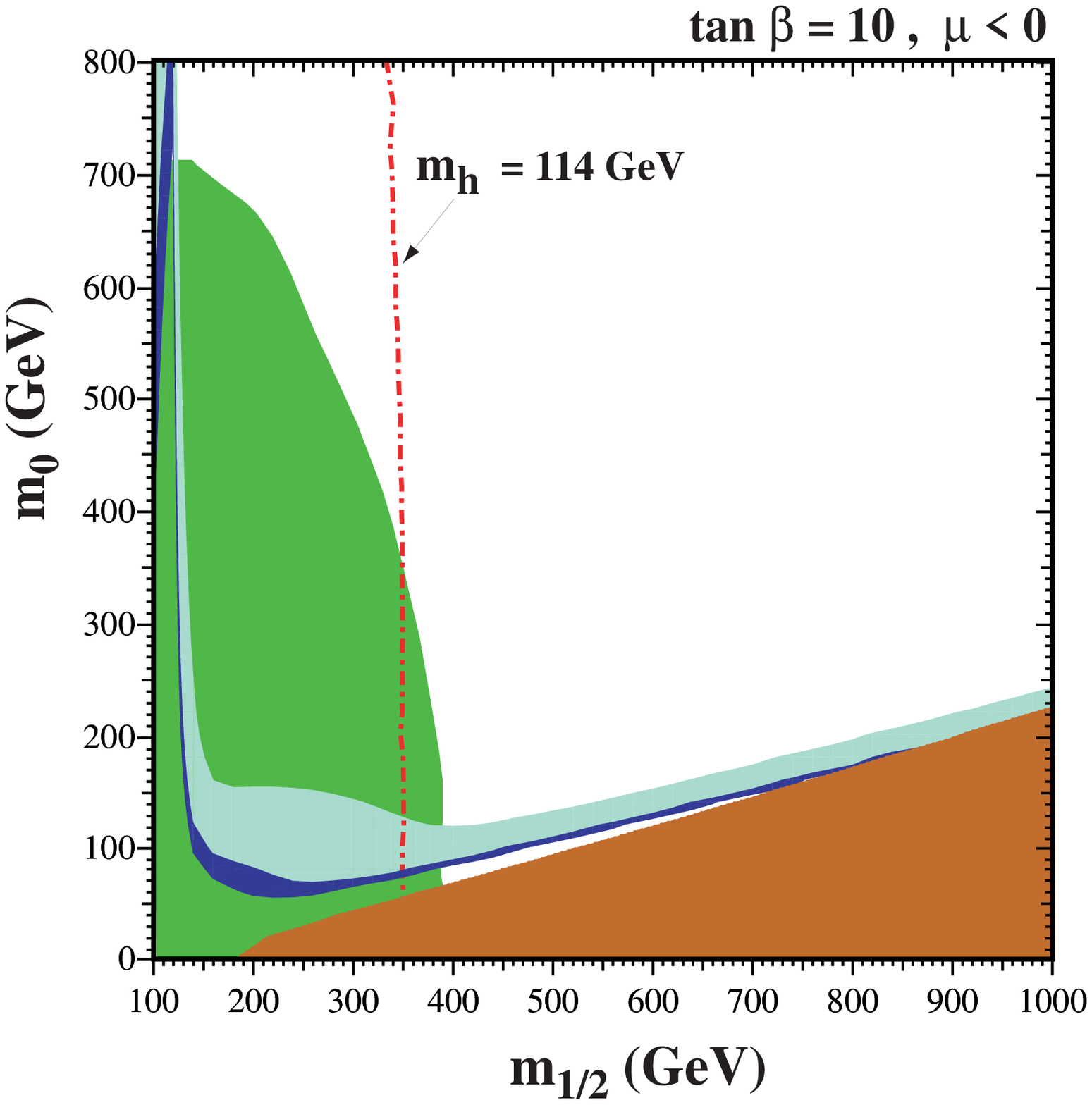}
}
\caption{\it
The $(m_{1/2}, m_0)$ planes for (left panel) $\tan\beta = 10, \mu > 0$, 
and (right panel) $\tan\beta = 10, \mu < 0$. In each panel, the
region allowed by the older cosmological constraint $0.1 \le \Omega_\chi
h^2 \le 0.3$ has light shading, and the region allowed by the newer WMAP
cosmological constraint $0.094 \le \Omega_\chi h^2 \le 0.129$ has very
dark shading. The region with dark (red) shading is disallowed 
because there the lightest supersymmetric particle would be charged.
The regions excluded by $b \rightarrow s \gamma$ have medium
(green) shading, and those in panels (a,d) that are favoured by $g_\mu -
2$ at the 2-$\sigma$ level have medium (pink) shading. LEP constraints on 
the Higgs and supersymmetric particle masses are also 
shown (Ellis {\it et al.} 2003$a$).
}
\vspace*{0.5cm}
\label{fig:EOSS}
\end{figure}

When considering the experimental, cosmological and theoretical
constraints on the MSSM, it is common to assume that all the unseen spin-0
supersymmetric particles have some universal mass $m_0$ at some GUT input
scale, and similarly for the unseen fermion masses $m_{1/2}$. These two
parameters of this constrained MSSM (CMSSM) are restricted by the absences
of supersymmetric particles at LEP: $m_{\chi^\pm} \gappeq 103$~GeV,
$m_{\tilde e} \gappeq 99$~GeV, and at the Fermilab
Tevatron collider. They are also restricted indirectly by the absence of a
Higgs boson at LEP: $m_h > 114.4$~GeV, and by the fact that $b \to s
\gamma$ decay is consistent with the Standard Model, and potentially by
the BNL measurement of $g_\mu - 2$, as seen in 
Fig.~\ref{fig:EOSS} (Ellis {\it et al.} 2003$a$; Lahanas \& Naopoulos 
2003).

As shown there, the CMSSM parameter space is also restricted by
cosmological bounds on the amount of cold dark matter, $\Omega_{CDM} h^2$.  
Since $\rho_\chi = m_\chi n_\chi$, and the relic number density $n_\chi
\propto 1 / \sigma_{ann} (\chi \chi \to ...)$ where $\sigma_{ann} (\chi
\chi \to ...) \propto 1 /m^2$, the relic density generically increases
with increasing sparticle masses. For some time, the conservative upper
limit on $\Omega_{CDM} h^2$ has been $< 0.3$ (Lahanas {\it et al.} 2000, 
2001a, 2001b; Barger \& Kao 2001; Arnowitt \& Dutta 2002), but the recent
WMAP data allow this to be reduced to $< 0.129$ at the 2-$\sigma$ level.
As seen in Fig.~\ref{fig:EOSS}, this improved upper limit significantly
improves the cosmological upper limit on the sparticle mass
scale (Ellis {\it et al.} 2003$a$, Lahanas \& Nanopoulos 2003). If there 
are other important components 
of the cold
dark matter, the CMSSM parameters could lie below the dark (blue) strips
in Fig.~\ref{fig:EOSS}.

In order to facilitate discussion of the physics reaches of different
accelerators and strategies for detecting dark matter, it is convenient to
focus on a limited number of benchmark scenarios that illustrate the
various different supersymmetric possibilities (Battaglia {\it et al.}
2001). In many of these scenarios, supersymmetry would be very easy to
observe at the LHC, and many different types of supersymmetric particle
might be discovered. However, searches for astrophysical dark matter may
be competitive for some scenarios (Ellis {\it et al.} 2001).

\begin{figure}[htb]
\centerline{\epsfxsize = 0.5\textwidth \epsffile{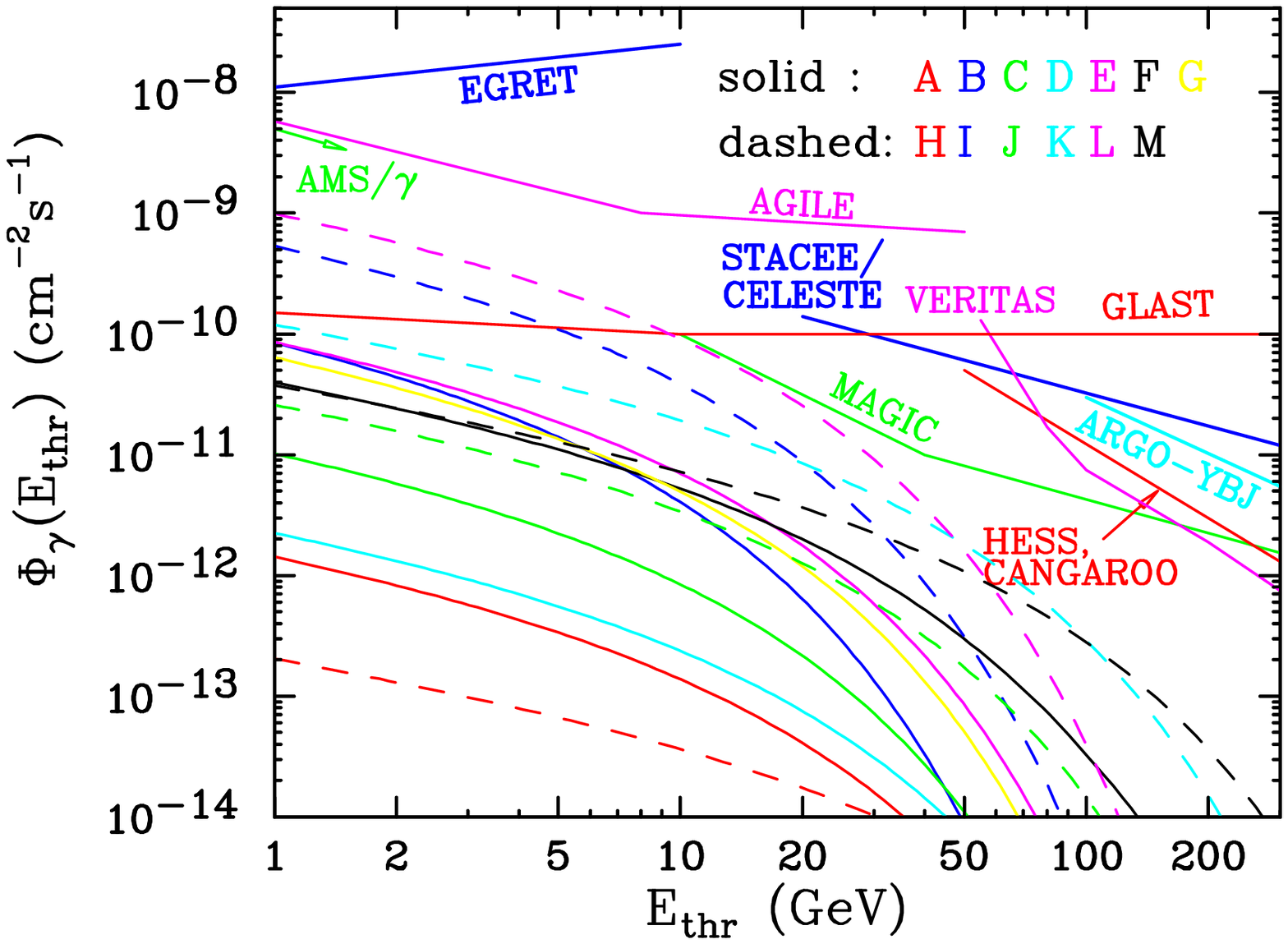}
\hfill \epsfxsize = 0.5\textwidth \epsffile{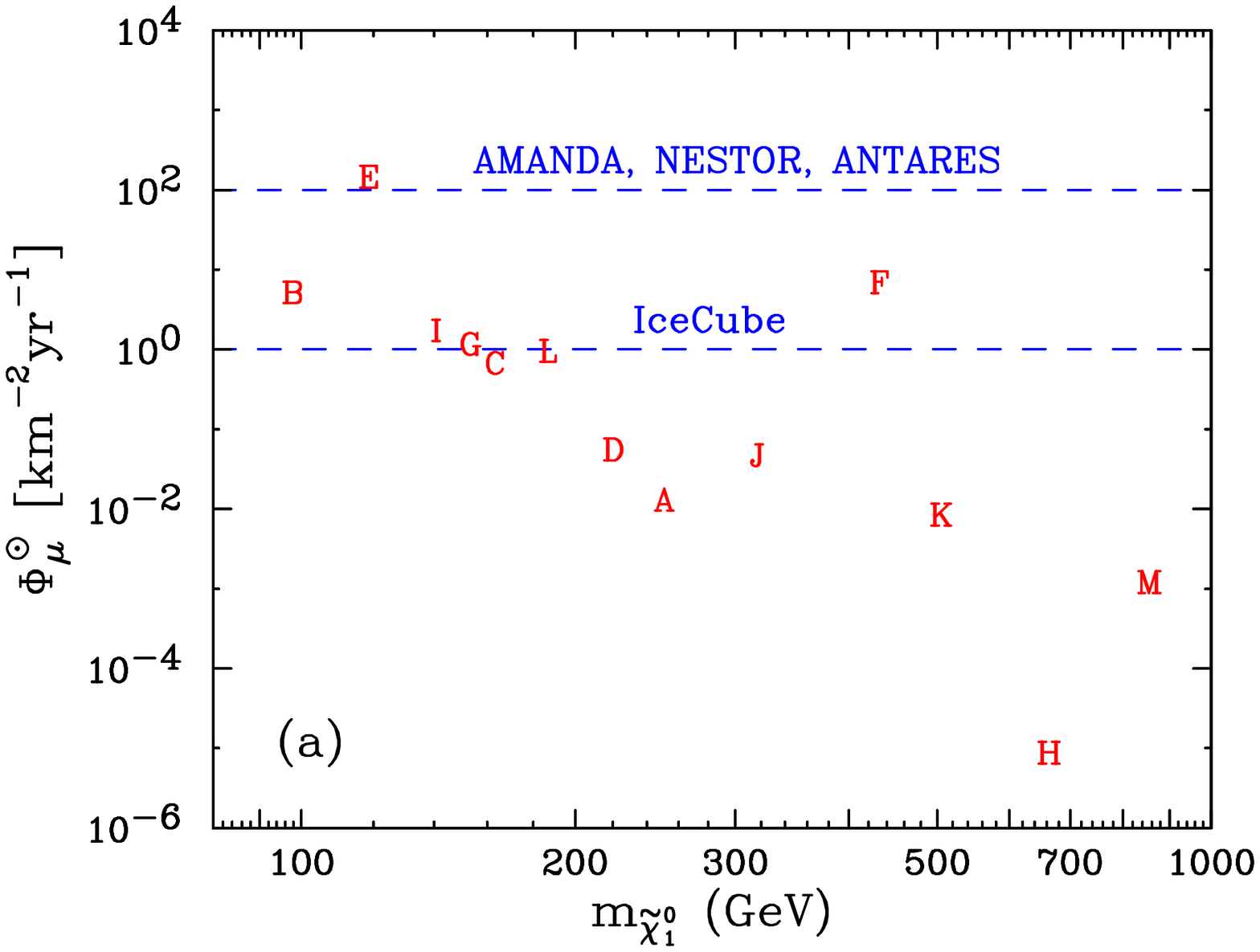}
}
\caption{\it
Left panel: Spectra of photons from the annihilations of
dark matter particles in the core of our galaxy, in different benchmark 
supersymmetric models (Ellis {\it et al.} 2001). Right panel: Signals for 
muons produced by energetic neutrinos originating from annihilations of dark 
matter particles in the core of the Sun, in the same benchmark    
supersymmetric models (Ellis {\it et al.} 2001).} 
\vspace*{0.5cm}
\label{fig:RS13}
\end{figure}

One strategy is to look for relic annihilations out in the galactic halo,
which might produce detectable antiprotons or positrons in the cosmic rays
(Silk \& Srednicki 1984). As discussed here by Carr (2003), both the
PAMELA (Pearce {\it et al.} 2002) and AMS (Aguilar {\it et al.} 2002)  
space experiments will be looking for these signals, though the rates are
not very promising in the benchmark scenarios we studied (Ellis {\it et
al.} 2001). Alternatively, one might look for annihilations in the core of
our galaxy, which might produce detectable gamma rays. As seen in the left
panel of Fig.~\ref{fig:RS13}, this may be possible in certain benchmark
scenarios (Ellis {\it et al.} 2001), although the rate is rather uncertain
because of the unknown enhancement of relic particles in our galactic
core. A third strategy is to look for annihilations inside the Sun or
Earth (Silk {\it et al.} 1985), where the local density of relic particles
is enhanced in a calculable way by scattering off matter, which causes
them to lose energy and become gravitationally bound. The signature would
then be energetic neutrinos that might produce detectable muons. As also
discussed here by Carr (2003), several underwater and ice experiments are
underway or planned to look for this signature, and this strategy looks
promising for several benchmark scenarios, as seen in the right panel of
Fig.~\ref{fig:RS13}~\footnote{It will be interesting to have such neutrino
telescopes in different hemispheres, which will be able to scan different
regions of the sky for astrophysical high-energy neutrino sources.}.

\begin{figure}[htb]
\centerline{\epsfxsize = 0.7\textwidth \epsffile{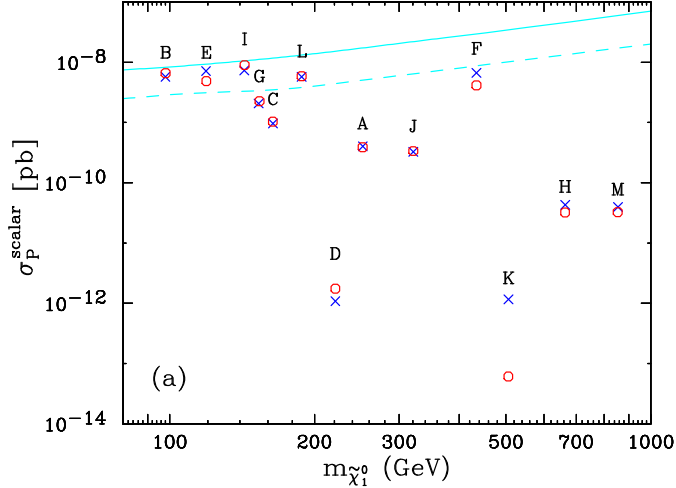}
}
\caption{\it
Rates calculated for the spin-independent elastic scattering of dark 
matter particles off protons in the same benchmark    
supersymmetric models (Ellis {\it et al.} 2001) as in 
Fig.~\ref{fig:RS13}.}
\vspace*{0.5cm}
\label{fig:RS11}
\end{figure}

The most satisfactory way to look for supersymmetric relic particles is
directly via their scattering on nuclei in a low-background laboratory
experiment Goodman \& Witten 1985). There are two types of scattering
matrix elements, spin-independent - which are normally dominant for
heavier nuclei, and spin-dependent - which could be interesting for
lighter elements such as fluorine. The best experimental sensitivities so
far are for spin-independent scattering, and one experiment has claimed a
positive signal (Bernabei {\it et al.} 1998). However, this has not been
confirmed by a number of other experiments, as discussed here by Kraus
(2003) and Smith (2003). In the benchmark scenarios the rates are
considerably below the present experimental sensitivities, but there are
prospects for improving the sensitivity into the interesting range, as
also discussed by Kraus (2003) and Smith (2003), as seen in
Fig.~\ref{fig:RS11}.

Overall, the searches for astrophysical supersymmetric dark matter have
discovery prospects (Ellis {\it et al.} 2001) that are comparable with 
those of the LHC (Battaglia {\it et al.} 2001).

\subsection{Superheavy Dark Matter}

\begin{figure}[htb]
\centerline{\epsfxsize = 0.7\textwidth \epsffile{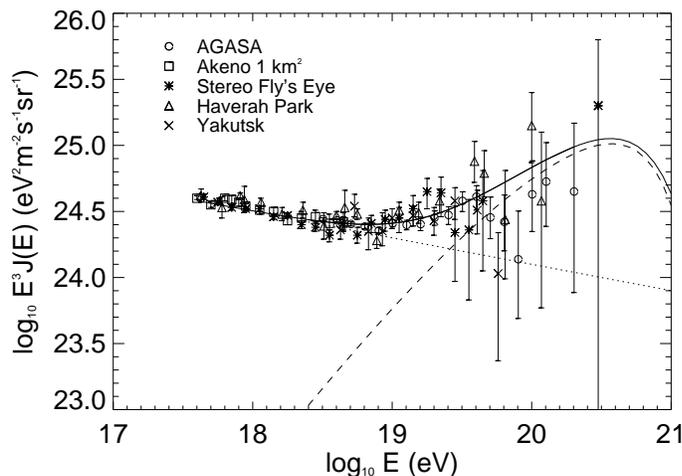}
}
\caption{\it
Calculation of the spectrum of ultra-high-energy cosmic rays that 
might be produced by the decays (Sarkar 2002) of metastable superheavy 
particles.} 
\vspace*{0.5cm}
\label{fig:RS19}
\end{figure}

As discussed here by Kolb (2003), it has recently been realized that
interesting amounts of superheavy particles with masses $\sim 10^{14 \pm
5}$~GeV might have been produced non-thermally in the very early Universe,
either via pre- and reheating following inflation, or during bubble
collisions, or by gravitational effects in the expanding Universe. If any
of these superheavy particles were metastable they might be able, as seen
in Fig.~\ref{fig:RS19}, to explain the ultra-high-energy cosmic rays that
seem (Takeda {\it et al.} 2002; Abu-Zayyad {\it et al.} 2002) to appear
beyond the Greisen-Zatsepin-Kuzmin (GZK) cutoff (Greisen 1966; Zatsepin \&
Kuzmin 1966).  Such models have to face some challenges, notably from
upper limits on the fractions of gamma rays at ultra-high energies, but
might exhibit distinctive signatures such as a galactic anisotropy (Sarkar
2002).  Examples of such superheavy particles are the cryptons found in
some models derived from string theory, which naturally have masses $\sim
10^{12 \pm 2}$~GeV and are metastable (like protons), decaying via
higher-order multiparticle interactions (Ellis {\it et al.} 1990; Benakli
{\it et al.} 1999). The Pierre Auger experiment (Cronin {\it et al.} 2002)
will be able to tell us whether ultra-high-energy cosmic rays really exist
beyond the GZK cutoff, and, if so, whether they are due to some such
exotic top-down mechanism, or whether they have some bottom-up
astrophysical origin. Following Auger, there are ideas for space
experiments such as EUSO (Petrolini {\it et al.} 2002) that could have
even better sensitivities to ultra-high-energy cosmic rays.

\section{Calculate it!}

Now that we have a good idea of the matter and energy content of the
Universe, and some prospects for detecting it, the next task is to
calculate it from first principles on the basis of microphysics and
laboratory data.

$\bullet$ $\Omega_b$: As Sakharov taught us (Sakharov 1967), baryogenesis
requires the violation of charge conjugation C and its combination CP with
parity, interactions that violate baryon number B, and a departure from
thermal equilibrium. The first two have been observed for quarks, and are
expected within the Standard Model. B violation is also expected in the
Standard Model, at the non-perturbative level. One might therefore wonder
whether the observed cosmological baryon asymmetry could have been
generated by the Standard Model alone, but the answer seems to be no
(Gavela {\it et al.} 1994). However, it might be possible in the MSSM, if
it contains additional sources of CP violation beyond the Standard Model
(Carena {\it et al.} 2003). An attractive alternative is leptogenesis
(Fukugita \& Yanagida 1986), according to which first the decays of heavy
singlet neutrinos create a CP-violating asymmetry $\Delta L \ne 0$, and
then this is partially converted into a baryon asymmetry by
non-perturbative weak interactions.

At the one-loop level, the asymmetry in the decays of one heavy singlet 
neutrino $N_i$ due to exchanges of another one, $N_j$, is
\begin{equation}
\epsilon_{ij} = {1 \over 8 \pi} {1 \over ( Y_\nu Y^\dagger_\nu)_{ii}}
{\rm Im} \left( (Y_\nu Y^\dagger_\nu)_{ij} \right)^2 f \left( {M_j \over 
M_i} \right),
\label{epsilon}
\end{equation}
where $Y_\nu$ is a matrix of Yukawa couplings between heavy singlet and 
light doublet neutrinos. The expression (\ref{epsilon}) involves a sum 
over the light leptons, and hence is independent of the CP-violating MNS 
phase $\delta$ and the Majorana phases $\phi_{1,2}$. Instead, it is 
controlled by extra phase parameters that are not directly accessible to 
low-energy experiments.

\begin{figure}[htb]
\centerline{\epsfxsize = 0.5\textwidth \epsffile{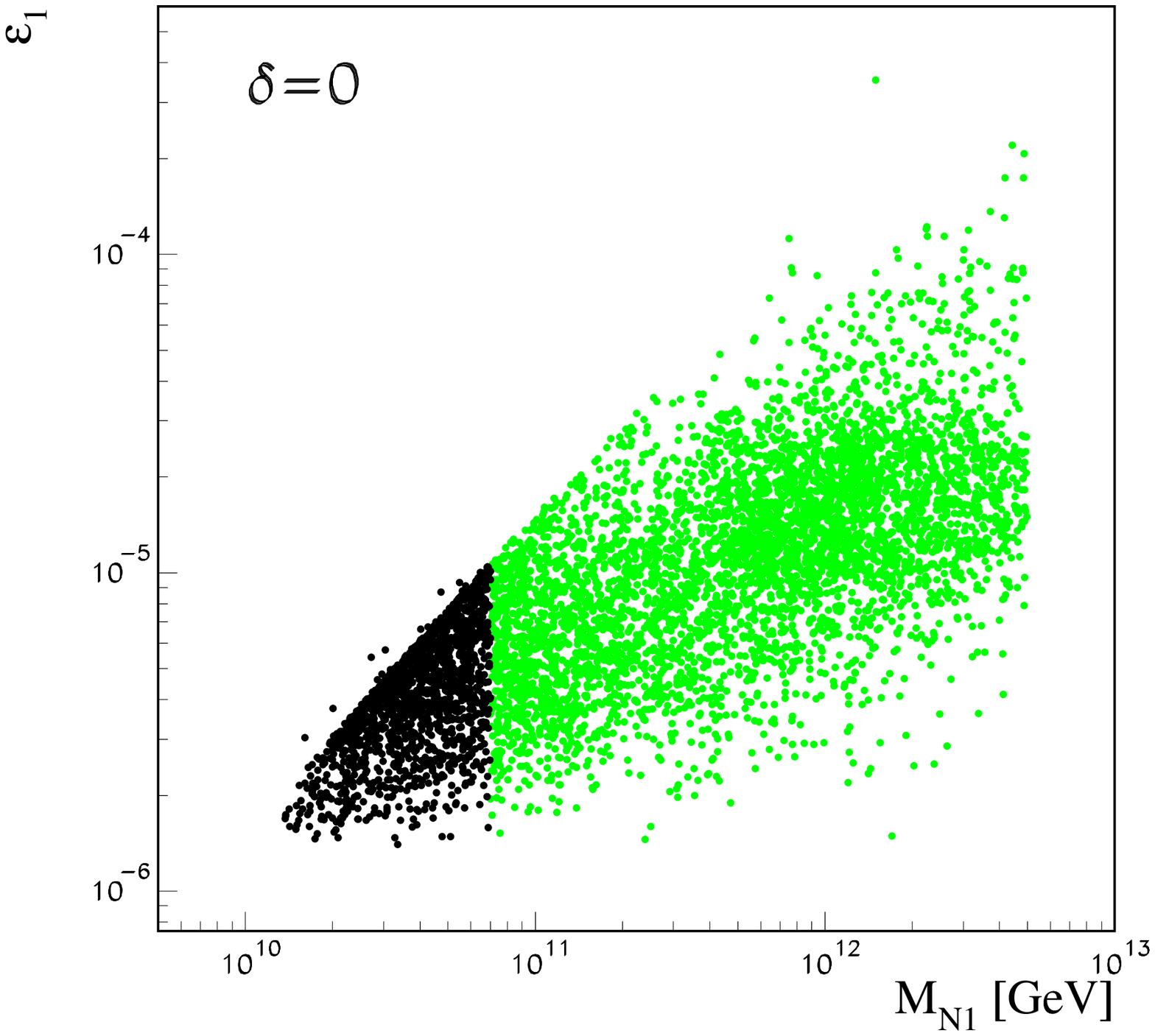}
\hfill \epsfxsize = 0.5\textwidth \epsffile{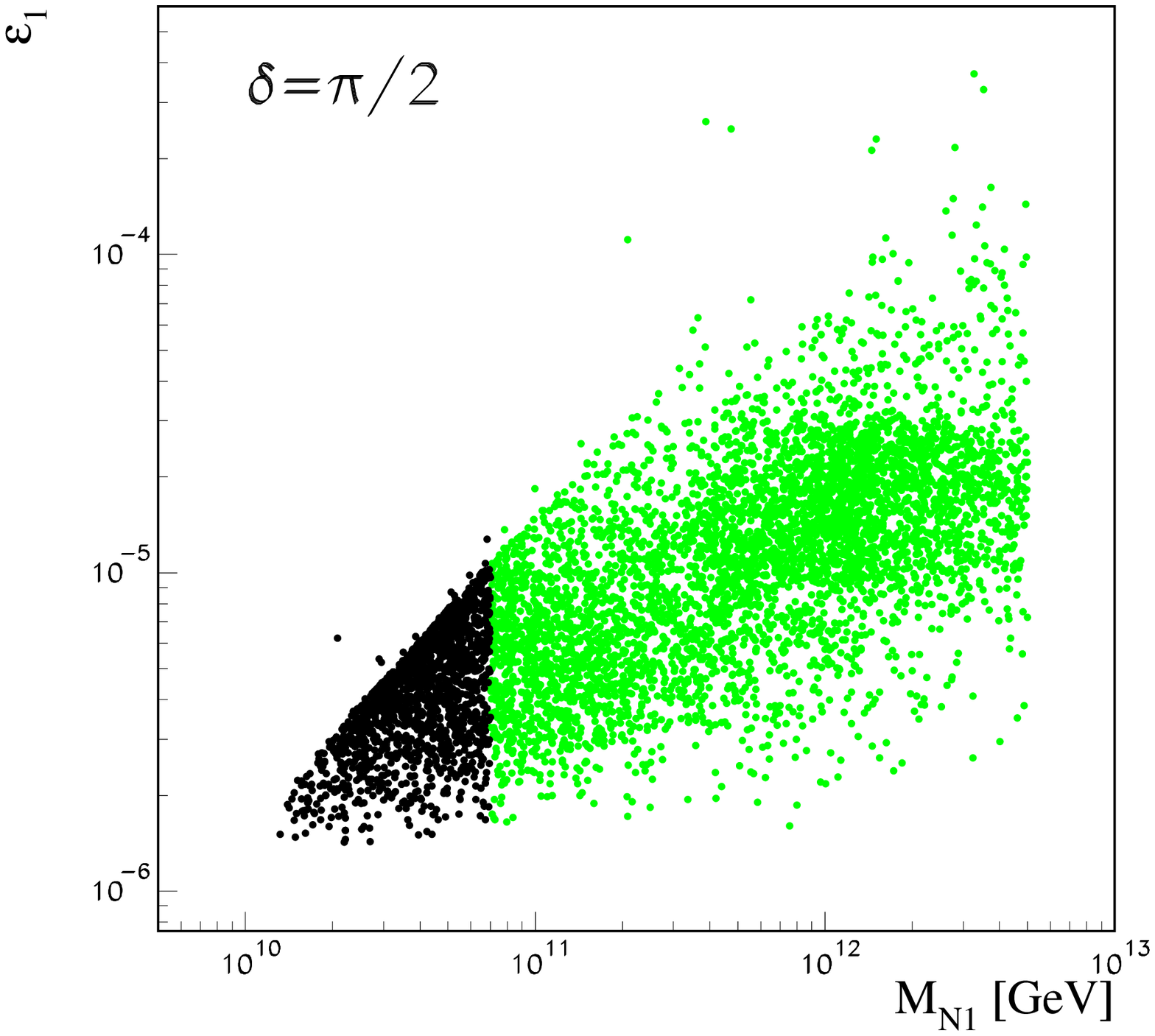}}
\caption{\it
Comparison of the CP-violating asymmetries in the decays of heavy singlet 
neutrinos giving rise to the cosmological baryon asymmetry via 
leptogenesis (left panel) without and (right panel) with maximal 
CP violation in neutrino oscillations (Ellis \& Raidal 2002). They are 
indistinguishable.} 
\vspace*{0.5cm} \label{fig:RS14}
\end{figure}

This leptogenesis scenario produces effortlessly a baryon-to-photon ratio
$Y_B$ of the right order of magnitude. However, as seen in
Fig.~\ref{fig:RS14}, the CP-violating decay asymmetry (\ref{epsilon}) is
explicitly independent of $\delta$ (Ellis \& Raidal 2002). On the other
hand, other observables such as the charged-lepton-flavour-violating
decays $\mu \to e \gamma$ and $\tau \to \mu \gamma$ may cast some indirect
light on the mechanism of leptogenesis (Ellis {\it et al.} 2002$a$, $b$).
Predictions for these decays may be refined if one makes extra hypotheses.

One possibility is that the inflaton might be a heavy singlet sneutrino
(Murayama {\it et al.} 1993, 1994).  This would require a mass $\simeq 2
\times 10^{13}$~GeV, which is well within the range favoured by seesaw
models. The sneutrino inflaton model predicts (Ellis {\it et al.} 2003$b$)
values of the spectral index of scalar perturbations, the fraction of
tensor perturbations and other CMB observables that are consistent with
the WMAP data (Peiris {\it et al.} 2003) . Moreover, this model predicts
a branching ratio for $\mu \to e \gamma$ within a couple of orders of
magnitude of the present experimental upper limit.

$\bullet$ $\Omega_{CDM}$: The relic density of supersymmetric dark matter
is calculable in terms of supersymmetric particle masses and Standard
Model parameters. The sensitivity to these parameters is quite small in
generic regions, but may be larger in some exceptional regions
corresponding to `tails' of the MSSM parameter space (Ellis \& Olive
2001). At least away from these regions, data from the LHC on
supersymmetric parameters should enable the cold dark matter density to be
calculated quite reliably.

\begin{figure}[htb]
\centerline{\epsfxsize = 0.8\textwidth \epsffile{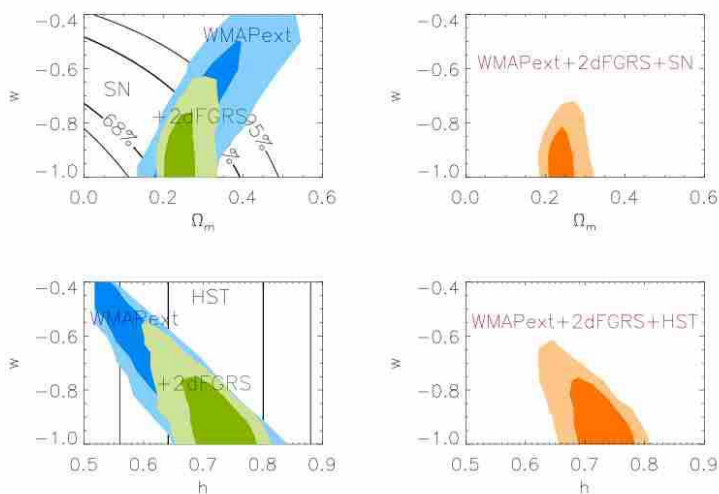}}
\caption{\it
Constraints on the equation of state of the dark energy from WMAP and 
other CMB data (WMAPext) combined with data on high-redshift supernovae, 
from the 2dF galaxy redshift survey and from the Hubble Space 
Telescope (Spergel {\it et al.} 2003).} 
\vspace*{0.5cm}
\label{fig:RS4}
\end{figure}

$\bullet$ $\Omega_\Lambda$: The biggest challenge may be the cosmological
vacuum energy.  For a long time, theorists tried to find reasons why the
cosmological constant should vanish, but no convincing symmetry to
guarantee this was ever found. Now cosmologists tell us that the vacuum
energy actually does not vanish. Perhaps theorists' previous failure
should be reinterpreted as a success? If the vacuum energy is indeed a
constant, the hope is that it could be calculated from first principles in
string or M theory. Alternatively, as argued here by Steinhardt 2003,
perhaps the vacuum energy is presently relaxing towards zero, as in
quintessence models (Maor {\it et al.} 2002).  Such models are getting to
be quite strongly constrained by the cosmological data, in particular
those from high-redshift supernovae and WMAP (Spergel {\it et al.} 2003)  
as seen in Fig.~\ref{fig:RS4}, and it seems that the quintessence equation
of state must be quite similar to that of a true cosmological constant
(Spergel {\it et al.} 2003). Either way, the vacuum energy is a
fascinating discovery that provides an exciting new opportunity for
theoretical physics.

\section{Test it!}

Laboratory experiments have explored the energy range up to about 100~GeV,
and quantum gravity must become important around the Planck energy $\sim
10^{19}$~GeV: where will new physics appear in this vast energy range?
There are reasons to think that {\it the origin of particle masses} will
be found at some energy $\lappeq 10^3$~GeV. If they are indeed due to some
elementary scalar Higgs field, this will provide a prototype for the
inflaton. If the Higgs is accompanied by supersymmetry, this may provide
the cold dark matter that fills the Universe. Some circumstantial evidence
for supersymmetry may be provided by the anomalous magnetic moment of the
muon and by measurements of the electroweak mixing angle $\theta_W$, in
the framework of {\it grand unified theories}. These would operate at
energy scales $\sim 10^{16}$~GeV, far beyond the direct reach of
accelerators. The first hints in favour of such theories have already been
provided by experiments on astrophysical (solar and atmospheric)  
neutrinos, and cosmology may provide the best probes of grand unified
theories, e.g., via inflation and/or super-heavy relic particles, which
might be responsible for the ultra-high-energy cosmic rays. Cosmology may
also be providing our first information about {\it quantum gravity}, in
the form of the vacuum energy.

The LHC will extend the direct exploration of the energy frontier up to
$\sim 10^3$~GeV. However, as these examples indicate, the unparallelled
energies attained in the early Universe and in some astrophysical sources
may provide the most direct tests of many ideas for physics beyond the
Standard Model. The continuing dialogue between the two may tell us the 
origins of dark matter and dark energy.

\end{document}